# A general thermodynamically consistent phase-field-micromechanics model of sintering with coupled diffusion and grain motion

Qingcheng Yang[1,2,*], Arkadz Kirshtein[3]

[1] Shanghai Key Laboratory of Mechanics in Energy Engineering, School of Mechanics and Engineering Science, Shanghai Institute of Applied Mathematics and Mechanics, Shanghai University, Shanghai 200072, China

[2] Shanghai Institute of Aircraft Mechanics and Control, Shanghai 200072, China

[3] Department of Mathematics & Statistics, Texas A&M University-Corpus Christi, Corpus Christi, TX 78412, United States

[*] Corresponding author, Email: qyang@shu.edu.cn, Tel: 86-18116136627.

# Abstract

Sintering is a pivotal technology for processing ceramic and metallic powders into solid objects. A profound understanding of microstructure evolution during sintering is essential for manufacturing products with tailored properties. While various phase-field models have been proposed to simulate microstructure evolution in solid-state sintering, correctly incorporating the crucial densification mechanism, particularly grain motion, remains a challenge. The fundamental obstacle lies in the ad hoc treatment of the micromechanics of grain motion, where the thermodynamical driving force cannot be derived from the system's free energy. This work presents a novel phase-field-micromechanics model for sintering (PFMMS) that addresses this challenge. We define a unified energy law, under which the governing equations for microstructure evolution in sintering are derived using variational principles. Our approach ensures thermodynamic consistency, with the driving force for grain motion derived from the system's free energy. Consequently, the proposed PFMMS guarantees the evolution of microstructure in a direction that reduces the system's energy and eliminates non-densifying phenomena. We rigorously validate PFMMS against recent benchmarks of theoretical and numerical analysis. It is found that PFMMS captures intrinsic stress distribution along and beyond grain boundaries, exhibits system-size-independent shrinkage strain, and maintains thermodynamic equilibrium states. These features are fundamental requirements for a physically consistent sintering model.

# 1. Introduction

Sintering stands as a fundamental manufacturing technique in materials processing, characterized by the consolidation of powdered materials through the application of heat. Its significance reverberates across diverse industries, from ceramics to metallurgy and beyond. By enabling the formation of dense, intricate structures, solid-state sintering finds application in crafting advanced ceramics for electronics, robust metal components for aerospace, and biomedical implants for healthcare.

In solid-state sintering, the driving force for the densification of powdered materials is to reduce the total sum of surface and grain boundary energies[1, 2]. The energy reduction is achieved through different mass transport mechanisms including surface diffusion, grain boundary diffusion, volume diffusion, vapor transport (evaporation-condensation), and grain motion, often referred to as rigid-body motion[3-7], as shown in Fig.1. The cooperative mass transport mechanisms complicate the microstructure evolution. Since the microstructure resulting from the sintering determines the properties of sintered products, it is essential to have a deep understanding of microstructure evolution to advance sintering and manufacture products with tailored properties. As such, numerous efforts have been explored along this direction to deepen our understanding, including analytical approach[4, 8-15], discrete element method[16-18], kinetic Monte Carlo (kMC) approach[19-22], and phase-field method [5-7, 23-46]. More discussions on modeling of sintering may be found in [2, 4, 32, 36, 47-49].

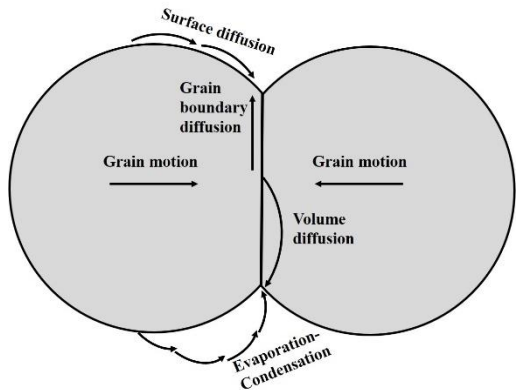

Figure 1: Schematic illustration of possible mass transport mechanisms in solid-state sintering.

The analytical approach is based on specifically idealized grain geometries to analyze sintering neck formation and densification, and is often employed for verifying advanced models. The discrete element method represents the motion of spherical grains interacting with their neighbors through appropriate sintering laws (normally derived from micromechanics analysis). The kinetic Monte Carlo model describes individual grains and pores using digitized discrete spatial points and the microstructure evolution is achieved by the Metropolis algorithm to switch the spatial sites, either representing a grain or pore.

Besides the aforementioned approaches, the phase-field method has been proven in the past decades as a powerful means to predict the microstructure evolution under different materials processes and is now widely employed in studying microstructure evolution in sintering [7, 32, 33, 36]. The development of phase-field modeling of solid-state sintering may be along the following directions:

(1) Thermodynamically, different phase-field energy functionals have been proposed. Existing works along this direction may be classified into two groups: the free-energy-based models considering only surface and grain boundaries energy [7, 23, 50, 51] and the grand-potential-based category [32, 36].

In the free-energy-based group, the fundamental idea is to express surface and grain boundary energies in terms of the introduced phase-field variables. A pioneering phase-field energy functional was proposed in [7], which is widely employed and extended to investigate microstructure evolution under different sintering conditions [27, 28, 30, 31, 34, 42, 52-56]; Recently, the pioneering phase-field energy functional in [7] was analyzed and was found to lack interfacial consistency and may introduce artificial voids at grain boundary junctions [50]. And a new free-energy-based phase-field functional was proposed in [50] to avoid such non-physical voids.

In the grand-potential-based group, the energy of the powdered system consists of surface energy, grain boundary energy, and grand potential contributions from the respective solid and vapor phases. Two representative models have been introduced recently in [32] and [36], and they differ from each other by employing different formulations of interfacial energies, with one using multiwell-type energy contribution [36] and the other using the obstacle-type potential energy [32].

(2) Kinetically, different mass transport mechanisms are considered. Existing works along this direction may be grouped into grain-motion-independent group [24, 27, 29, 32, 34, 36, 50, 53, 57] and grain-motion-dependent category [5, 6, 39, 40, 43, 51, 55, 56].

In the grain-motion-independent group, only different diffusion mechanisms including surface diffusion, grain boundary diffusion, volume diffusion, and evaporation and condensation, are considered as mass transport paths. Recently, it was recognized that the grain-motion-independent models introduce less pronounced shrinkage compared to the grain-motion-dependent models and have a system-size-dependent shrinkage rate, which is unphysical [5]. The significance of grain motion contribution to densification was also emphasized in recent analysis work in [4, 12].

In the grain-motion-dependent category, the basic idea is to augment the standard phase-field equations with an advection term describing the effect of mass transport due to grain motion, in addition to the different diffusion mechanisms. A pioneering work was the ad hoc rigid-body motion formula proposed in [7], which is commonly employed and further developed in the community. Recently it was found that the rigid-body motion formula in [7] may introduce a non-densifying phenomenon and increase the system's free energy after a certain densifying point [5]. This has motivated further improvements developed in [5, 39, 43, 44]. Furthermore, its coupling with the discrete element method and solid mechanics are also investigated in [31, 51, 56].

However, although the significance of grain motion's contribution to sintering shrinkage has been recognized and emphasized [4, 5] and numerous efforts have been made to account for this contribution, it is still a challenge to incorporate grain motion correctly in existing approaches, as recently mentioned in [4]. The fundamental reason is that the micromechanics of grain motion during sintering is either not considered or treated in an ad hoc manner [4]. As such, the driving force for grain motion in sintering cannot be derived from the system's free energy by the variational principle. Thus, the thermodynamical consistency is not guaranteed and a non-densification phenomenon may arise [5].

As an effort to address this fundamental and long-existing challenge, a general, thermodynamically consistent, and simple phase-field-micromechanics model of sintering (PFMMS) is proposed in this work. The widely employed rigid body motion assumption is relaxed and grain particles are allowed to have possible mechanical deformation through surface and grain-boundary tensions in PFMMS. The generality of PFMMS lies in that the proposed framework may apply to different phase-field energy functionals as the system's energy may be expressed in different mathematical forms. The thermodynamical consistency of PFMMS lies in that the phase-field equations depicting microstructure evolution and the equation governing the micromechanics of grain motion are all derived from the system's free energy under a unified energy law and that mass conservation and balance of linear momentum of the system are both satisfied. The simplicity of PFMMS lies in that the incorporation of grain motion is handled naturally mechanics-wise such that it does not require extra parameters and filtering functions as are often needed in existing works.

As such, the proposed PFMMS guarantees that the evolution of microstructure is along a direction that reduces the system's energy monotonically and can depict the stress distribution both along grain boundaries and within the grain compacts. The proposed PFMMS is rigorously validated against benchmark examples from recent theoretical and numerical analysis, as suggested in [4, 5].

The remainder of the manuscript is organized as follows. Section 2 introduces the general framework of the phase-field-micromechanics model of sintering (PFMMS), emphasizing its thermodynamic consistency and applicability to various free energy functionals. Furthermore, as a specific example, the recently proposed phase-field energy functional is selected to address artificial void generation at grain junctions. In Section 3, different benchmark examples from recent theoretical and numerical analysis are employed to rigorously validate PFMMS. Section 4 discusses the modeling of the micromechanics of grain motion in sintering. Finally, Section 5 concludes the performance of the proposed PFMMS and outlines the future research directions.

## 2. The general framework

This section presents the general framework of the proposed phase-field-micromechanics model of sintering (PFMMS) using the mixture theory. Then a specific phase-field energy functional is selected as a representative example under the proposed framework. The applicability to other existing phase-fiend energy functionals of sintering is also discussed.

2.1 Conservation of mass

In solid-state sintering, the powdered system is treated as a two-phase mixture, i.e., vapor-grain mixture. Let $\rho_g^0$ and $\rho_v^0$ represent the initial densities of the respective grain phase and vapor at a spatial point before the mixture; let $\rho_g$ and $\rho_v$ be the densities, $V_g$ and $V_v$ be the volumes of the respective phases at a spatial point in the current mixture system. According to the mixture theory [58-61], we have:

$$C_g = \frac{V_g}{V_g + V_v} = \frac{\rho_g}{\rho_g^0} \tag{1}$$

$$C_v = \frac{V_v}{V_g + V_v} = \frac{\rho_v}{\rho_v^0} \tag{2}$$

$$C_g + C_v = 1 \tag{3}$$

where $C_g$ and $C_v$ denote the respective volume fractions of the grain phase and vapor phase ranging between 0 and 1 after the mixture. According to the mass conservation law of each phase in the current configuration, we have:

$$\frac{\partial \rho_g}{\partial t} + \nabla \cdot (\dot{\mathbf{u}}_g \rho_g) = 0 \tag{4}$$

$$\frac{\partial \rho_v}{\partial t} + \nabla \cdot (\dot{\mathbf{u}}_v \rho_v) = 0 \tag{5}$$

where $\dot{\mathbf{u}}_g$ and $\dot{\mathbf{u}}_v$ represent the respective velocities of each phase in the mixture system. According to Eqs. (1)-(5), equivalently, we also have:

$$\frac{\partial C_g}{\partial t} + \nabla \cdot (\dot{\mathbf{u}}_g C_g) = 0 \tag{6}$$

$$\frac{\partial C_v}{\partial t} + \nabla \cdot (\dot{\mathbf{u}}_v C_v) = 0 \tag{7}$$

Adding Eqs. (6) and (7) up and using Eq. (3), we have:

$$\nabla \cdot \dot{\mathbf{u}} = 0 \tag{8}$$

where $\dot{\mathbf{u}}$ is defined as:

$$\dot{\mathbf{u}} = \dot{\mathbf{u}}_g C_p + \dot{\mathbf{u}}_v C_v \tag{9}$$

It is evident that $\dot{\mathbf{u}}$ represents the volume-averaged background velocity of the two-phase mixture system. It is noteworthy that the velocity $\dot{\mathbf{u}}_g$ consists of two independent contributions: the background velocity $\dot{\mathbf{u}}$ from grain motion and the drift velocity $\dot{\mathbf{u}}_g^*$ arising from diffusion:

$$\dot{\mathbf{u}}_g = \dot{\mathbf{u}} + \dot{\mathbf{u}}_g^* \tag{10}$$

To provide a clearer understanding of the physical interpretations of $\dot{\mathbf{u}}$ and $\dot{\mathbf{u}}_g^*$, we draw an analogy: $\dot{\mathbf{u}}$ can be likened to the mechanics velocity representing the movement of a river carrying salts, while $\dot{\mathbf{u}}_g^*$ is akin to the drift velocity describing the diffusion of salts within the river. Since $\dot{\mathbf{u}}$ arises from grain motion, it is required to satisfy the balance of linear momentum. In contrast, $\dot{\mathbf{u}}_g^*$, being a result of diffusion, is governed by Fick's diffusion law.

For simplicity, let $C$ denote $C_g$, the volume fraction of the grain phase in the mixture, by dropping off the subscript. According to Eqs. (6) and (10), we have:

$$\frac{\partial C}{\partial t} + \nabla \cdot (\dot{\mathbf{u}} C) + \nabla \cdot (\dot{\mathbf{u}}_g^* C) = 0 \tag{11}$$

As such, the flux of $C$ has contributions from both $\dot{\mathbf{u}}$ and $\dot{\mathbf{u}}_g^*$. Essentially, $C$ functions as a conventionally employed conservative phase-field variable ranging from 0 to 1 to differentiate the solid grain phase from the surrounding vapor phase.

## 2.2 The energy law

In addition to the introduced field variable $C$ in section 2.1, another set of field variables $\eta_i$ is needed to represent the volume fraction of the $i$th grain in the grain-grain mixture. As such, $\eta_i$ ranges from 0 to 1 and differentiates the $i$th grain from the rest with different grain orientations. Essentially, the introduced $C$ and $\eta_i$ function as the conventionally employed phase-field variables and collectively describe the microstructure features in the powdered system. The free energy functional $\mathcal{F}$ of the mixture may generally be expressed as a function of $C$ and $\eta_i$, as follows:

$$\mathcal{F} = \int_\Omega f\left(C, \eta_1, \eta_1, \ldots, \eta_{n_g}, \nabla C, \nabla \eta_1, \nabla \eta_2, \ldots, \nabla \eta_{n_g}\right) d\Omega \tag{12}$$

where $n_g$ denotes the total number of grains with different orientations. Physically, $\mathcal{F}$ represents the sum of surface energy and grain boundary energies within the mixture. Let $\mathcal{D}$ denote the total energy dissipation rate of the system, comprising three distinct contributions in sintering: $\mathcal{D}_C$ due to diffusion (grain-vapor interaction), $\mathcal{D}_\eta$ arising from grain boundary evolution (grain-grain interaction), and $\mathcal{D}_m$ originating from the mechanics of grain motion, characterized by the velocity $\dot{\mathbf{u}}$ as defined in section 2.1. Assuming negligible kinetic energy from grain motion (due to the slow nature of sintering) and negligible elastic energy, the energy law governing the powdered system can be defined as follows:

$$\frac{d\mathcal{F}}{dt} = -\mathcal{D} \tag{13}$$

$$\mathcal{D} = \mathcal{D}_C + \sum_i^{n_g} \mathcal{D}_{\eta_i} + \mathcal{D}_m \tag{14}$$

where $t$ is time. Note that the choices of $\mathcal{F}$, $\mathcal{D}_C$, $\mathcal{D}_{\eta_i}$, and $\mathcal{D}_m$ take into consideration all the physics of the powdered system and determine the governing equations for the phase-field variables $C$ and $\eta_i$, the mechanical velocity $\dot{\mathbf{u}}$ and the drift velocity $\dot{\mathbf{u}}_g^*$ through variational principle. In this work, $\mathcal{D}_C$ and $\mathcal{D}_{\eta_i}$ are chosen as follows:

$$\mathcal{D}_C = \int_\Omega \frac{1}{2} C^2 M^{-1} \left\|\dot{\mathbf{u}}_g^*\right\|^2 d\Omega \tag{15}$$

$$\mathcal{D}_{\eta_i} = \int_\Omega \frac{1}{2} L^{-1} \left\| \frac{\mathrm{D}\eta_i}{\mathrm{D}t} \right\|^2 \mathrm{d}\Omega \tag{16}$$

where $M$ is atomic mobility, $L$ is the grain boundary mobility; $\frac{\mathrm{D}\eta_i}{\mathrm{D}t}$ is the material derivative of $\eta_i$ defined in continuum mechanics and can be written in Eulerian description as:

$$\frac{\mathrm{D}\eta_i}{\mathrm{D}t} = \dot{\eta} + \dot{\mathbf{u}} \cdot \nabla \eta_i \tag{17}$$

It will be shown in Appendix A that with the chosen $\mathcal{D}_C$ and $\mathcal{D}_\eta$, the conventional phase-field equations, i.e., the Cahn-Hillard and Allen-Cahn equations, will be recovered by the variational principle. According to Ficks's law, the drift velocity $\dot{\mathbf{u}}_g^*$ is governed by diffusion flux equivalence as follows:

$$\dot{\mathbf{u}}_g^* C = -M \nabla \frac{\delta \mathcal{F}}{\delta C} \tag{18}$$

(derived from variations; See Appendix A for details). For mechanics dissipation $\mathcal{D}_m$ in this work, the solid grain phase is assumed to be viscous at elevated temperature [62] and $\mathcal{D}_m$ is chosen as:

$$\mathcal{D}_m = \int_\Omega \frac{1}{2} \mu^{eff} \| \nabla \dot{\mathbf{u}} + (\nabla \dot{\mathbf{u}})^\mathrm{T} \|^2 \mathrm{d}\Omega \tag{19}$$

where $\mu^{eff}$ is the effective viscosity of the mixture and is defined as:

$$\mu^{eff} = \big( \varphi + (1-\varphi) N(C) \big) \mu_g \tag{20}$$

where $\mu_g$ is the viscosity of the grain phase, $\varphi$ denotes the ratio of vapor viscosity $\mu_v$ to grain phase viscosity $\mu_g$, and $N(C)$ represents the interpolation function given as [50]:

$$N(C) = C^2 [1 + 2(1-C) + \epsilon(1-C)^2], \epsilon > 3 \tag{21}$$

It takes a value of 0 in the vapor phase and 1 in the grain phase. The constant $\epsilon$ is chosen to be 3.1 in this work to guarantee $N(C)$ has a convex shape in the vicinity of $C = 1$. It is noteworthy that the chosen mathematical forms of $\mathcal{D}_C$, $\mathcal{D}_\eta$, and $\mathcal{D}_m$ maintain the non-negativity in energy dissipation. Consequently, the energy law described in Eq. (13) ensures that the energy of the powdered system decreases at a rate determined by the respective dissipation physics as the microstructure evolves.

2.3 Governing equations

Given the energy law defined in section 2.2, the governing equations of the sintering system can be derived following the variational principle. The details of the derivations are given in Appendix A. The employed governing equations of the powdered mixture in PFMMS are summarized as follows:

$$\frac{\partial C}{\partial t} + \nabla \cdot (C\dot{\mathbf{u}}) = \nabla \cdot \left(M \nabla \frac{\delta \mathcal{F}}{\delta C}\right) \tag{22}$$

$$\frac{\partial \eta_i}{\partial t} + \dot{\mathbf{u}} \cdot \nabla \eta_i = -L \frac{\delta \mathcal{F}}{\delta \eta_i} \tag{23}$$

$$\nabla \cdot \boldsymbol{\sigma} + \mathbf{b} = \mathbf{0} \tag{24}$$

$$\nabla \cdot \dot{\mathbf{u}} = 0 \tag{25}$$

where the Cauchy stress tensor $\boldsymbol{\sigma}$ and the body force vector $\mathbf{b}$ are defined as:

$$\boldsymbol{\sigma} = \mu^{eff}(\nabla \dot{\mathbf{u}} + (\nabla \dot{\mathbf{u}})^T) + p\mathbf{I} \tag{26}$$

$$\mathbf{b} = -\left(\nabla \cdot \left(\frac{\partial f}{\partial \nabla C} \otimes \nabla C\right) + \sum_i^{N_p} \left[\nabla \cdot \left(\frac{\partial f}{\partial \nabla \eta_i} \otimes \nabla \eta_i\right)\right] - \nabla f(C, \nabla C, \eta_i, \nabla \eta_i)\right) \tag{27}$$

In Eq.(26), the term $p$ represents the pressure and $\mu^{eff}$ denotes the effective viscosity defined in Eq. (20). It is important to highlight that the body force vector $\mathbf{b}$ is intrinsic and consists of contributions from both surface tension and grain boundary tension. The atomic mobility $M$, which considers different diffusion paths, is defined as [63]:

$$M = D^{eff} \left(\frac{\partial^2 \mathcal{F}}{\partial C^2}\bigg|_{C=1}\right)^{-1} \tag{28}$$

$$D^{eff} = D^{sf} C^2 (1-C)^2 + 4D^{gb} N(C) \sum_i^{N_p} \sum_{j \neq i}^{N_p} \eta_i^2 \eta_j^2 + D^{vol} N(C) \tag{29}$$

where $D^{eff}$, $D^{sf}$, $D^{gb}$, $D^{vol}$ are the effective, surface, grain boundary, and volume diffusivities, respectively.

The combination of Eqs. (22) and (25) represents the mass conservation of the powdered system, as detailed in section 2.1. Eq. (23) governs the interaction between grains (the dynamics of grain boundary evolution), while Eq. (24) ensures the mechanical equilibrium of grain motion. These characteristics, along with the energy law defined in Eq. (13), demonstrate the thermodynamic consistency of PFMMS. The unknown fields in PFMMS are the phase-field variables $C$ and $\eta_i$, the pressure $p$, and the background velocity $\dot{\mathbf{u}}$, all of which can be determined by solving the coupled equations.

Note that in the proposed PFMMS framework, the phase-field energy functional $\mathcal{F}$ is presented in a generalized form. Consequently, the derived governing equations (Eqs. (22)-(25)) are also generalized and can be applied to different forms of $\mathcal{F}$, showcasing the versatility of the proposed PFMMS. As a representative example, the recently proposed $\mathcal{F}$ in [50] will be selected in Section 2.5 due to its ability to avoid artificial void generation at grain junctions.

2.4 Relationship to existing rigid body motion models

The mechanical velocity $\dot{\mathbf{u}}$ depicts grain motion and essentially produces a convection mass flux in addition to diffusion flux as in Eq. (22). In existing phase-field models, the convection velocity $\dot{\mathbf{u}}_{RB}$ describing grain motion is defined based on the well-known rigid body model proposed in [7]. The relationships between $\dot{\mathbf{u}}$ in the proposed PFMMS and $\dot{\mathbf{u}}_{RB}$ in the existing rigid body motion models [5-7, 26, 40] are discussed as follows:

(1) Micromechanics: the micromechanics aspect of the proposed PFMMS is characterized by the inclusion of $\dot{\mathbf{u}}$ to describe grain motion, necessitating adherence to mechanical equilibrium principles. The governing equation for $\dot{\mathbf{u}}$, denoted as Eq. (24), is derived through variation of the utilized free energy concerning grain trajectory, ensuring thermodynamic consistency and directing its evolution toward minimizing total energy, as evidenced by the energy law and forthcoming numerical demonstrations. Consequently, the sintering stress tensor information across the grain compact can be extracted following Eq. (26).

In contrast, conventional rigid body models employ $\dot{\mathbf{u}}_{RB}$ in an ad hoc manner, lacking a direct derivation from free energy. Notably, $\dot{\mathbf{u}}_{RB}$ behaves akin to a drifting velocity, its relationship to force defined through mobility [7]. This approach may not be able to provide comprehensive stress tensor information and may not lead to energy minimization, as observed in [5]. Furthermore, $\dot{\mathbf{u}}$ inherently encompasses all possible translational and rotational components, while $\dot{\mathbf{u}}_{RB}$ necessitates a more complicated technique of calculating individual translational and rotational components to yield the total velocity[7].

(2) Rigidity: The rigidity aspect of $\dot{\mathbf{u}}$ as a field variable is characterized by a unified equation governing its behavior across the entire powdered compact. Within individual grains, $\dot{\mathbf{u}}$ is not mandated to remain uniform, with its rigidity regulated by $\mu_g$, representing the viscosity of the grain phase. A higher $\mu_g$ value results in a more uniform $\dot{\mathbf{u}}$ within a grain. Conversely, $\dot{\mathbf{u}}_{RB}$ is defined grain-wise and must maintain uniformity within each grain. Consequently, $\dot{\mathbf{u}}$ exhibits less restrictiveness compared to $\dot{\mathbf{u}}_{RB}$.

(3) Parameter employment: In the calculation of $\dot{\mathbf{u}}$ using Eq. (24), only one parameter, namely the grain phase viscosity $\mu_g$, is required. It's important to note that $\mu_g$ is utilized to characterize the mechanical behavior of grains at elevated temperatures. Physically, $\mu_g$ depends on factors such as the diffusivity of the grain phase and the sintering temperature, as discussed in [4, 12]. Therefore, the simplicity of PFMMS lies in its use of no extra parameter to describe the micromechanics of grain motion. In contrast, $\dot{\mathbf{u}}_{RB}$ in existing rigid body models necessitates the use of multiple parameters and a grain-boundary filter function for determination. These parameters may include stiffness, equilibrium density at the grain boundary, threshold values in the selected filter function, and translational and rotational mobilities, which can be challenging to determine [5, 7].

(4) Compatibility to existing libraries and computational cost: In PFMMS, $\dot{\mathbf{u}}$ is governed by a unified partial differential equation, i.e., Eq. (24), facilitating its seamless integration with existing PDE-based libraries such as MOOSE and FEniCS. Moreover, Eq. (24) eliminates the need to establish neighboring relationships among grains, thereby reducing computational costs. On the contrary, calculating $\dot{\mathbf{u}}_{RB}$ involves evaluating several algebraic equations and necessitates searching for neighboring grains for each grain, which requires additional effort for integration with existing PDE-based libraries. This process can be costly, especially when defining grain neighboring relationships for large-scale simulations.

2.5 A specific case

Under the general framework of the proposed PFMMS, the phase-field free energy functional $\mathcal{F}$ proposed in [50] is employed to avoid non-physical void generation at grain junctions:

$$\mathcal{F} = \mathcal{F}_{sf} + \mathcal{F}_{gb} \tag{30}$$

$$\mathcal{F}_{sf} = \int_\Omega \left( \alpha C^2 (1-C)^2 + \frac{\kappa_C}{2} \|\nabla C\|^2 \right) d\Omega \tag{31}$$

$$\mathcal{F}_{gb} = \int_\Omega N(C) \left( \beta \left[ 1 - 4 \sum_{i=1}^{N_p} \eta_i^3 + 3 \left( \sum_{i=1}^{N_p} \eta_i^2 \right)^2 \right] + \frac{\kappa_\eta}{2} \sum_{i=1}^{N_p} \|\nabla \eta_i\|^2 \right) d\Omega \tag{32}$$

where $\mathcal{F}_{sf}$ and $\mathcal{F}_{gb}$ represent the respective surface energy and grain boundary energy; $\alpha$ and $\beta$ are model constants; $\kappa_C$ and $\kappa_\eta$ are the respective gradient coefficients. The relationship between the phase-field parameters and material properties is given as:

$$\begin{cases} \gamma^{sf} = \frac{\sqrt{2\kappa_C \alpha}}{6} \\ \gamma^{gb} = \frac{2}{\sqrt{3}}\sqrt{\beta \kappa_\eta} \\ \delta^{sf} = \sqrt{\frac{8\kappa_C}{\alpha}} \\ \delta^{gb} = \sqrt{\frac{4\kappa_\eta}{3\beta}} \end{cases} \text{or} \begin{cases} \alpha = 12\frac{\gamma^{sf}}{\delta^{sf}} \\ \beta = \frac{\gamma^{gb}}{\delta^{gb}} \\ \kappa_C = \frac{3}{2}\gamma^{sf}\delta^{sf} \\ \kappa_\eta = \frac{3}{4}\gamma^{gb}\delta^{gb} \end{cases} \quad (33)$$

where $\gamma^{sf}$, $\gamma^{gb}$ are the specific surface energy and specific grain boundary energy, respectively; $\delta^{sf}$ and $\delta^{gb}$ are the respective surface thickness and grain boundary thickness. The detailed derivation of the relationship map is given in [50]. In the chosen $\mathcal{F}$, it's worth noting that the representation of the $i$th grain involves the product $C\eta_i$, which differs from existing models that solely employ $\eta_i$.

## 3. Numerical examples

In this section, the phase-field energy functional $\mathcal{F}$ selected in Section 2.5 will be utilized in Eqs. (22)-(25) to validate the proposed PFMMS against recent benchmarks from both theoretical and numerical analyses related to fundamental aspects of sintering, as suggested in [4, 12]. The finite element method in space and the semi-implicit finite difference method in time are used to solve Eqs. (22)-(25) under no flux and stress-free boundary conditions. The detailed numerical scheme employed will be discussed in a separate publication.

3.1 Classical Two-particle model

The classical two equally-sized-particle model is utilized here for comparison with recent theoretical analysis in [12] as it enables a fundamental examination of sintering. It's important to note that the grain phase viscosity $\mu_g$ is not an independent parameter but is inversely proportional to the grain boundary diffusivity $D^{gb}$ [12], as $\mu_g$ is employed to describe the mechanical behavior of grains due to diffusion. Moreover, given that the assumption of rigid grain particles in the theoretical analysis and deformable grain particles in PFMMS and that the values of employed parameters in [12] are not given, the validation of PFMMS is limited to qualitative comparisons with the theoretical work.

3.1.1 Stress distribution

The stress distribution along the grain boundary is analyzed first since it may be the most fundamental quantity to validate as it drives the densification process and determines the rest sintering kinetics [12, 64]. The normalized parameters utilized in PFMMS are presented in Table 1. The simulation domain is set to be 100 by 80 and is spatially discretized by finite element with 450 elements along the horizontal direction and 360 elements along the vertical direction. Figure 2 depicts the distribution of dimensionless normal stress $\sigma_{11}$ along the grain boundary as predicted by the employed PFMMS. The plot reveals a parabolic distribution with maximal magnitudes observed at the ends of the grain boundary and minimal magnitude at the midpoint. It's worth noting that this parabolic stress distribution along the grain boundary aligns with the theoretical predictions outlined in [12].

Table 1 The normalized parameter employed in stress calculation in PFMMS. $\mu_g$ is not an independent parameter and is inverse propositional to $D^{gb}$. The viscosity ratio $\varphi$ is set to be $\varphi = 0.0001$ in this manuscript if not mentioned otherwise.

| $\alpha$ | $\beta$ | $\kappa_C$ | $\kappa_\eta$ | $D^{sf}$ | $D^{gb}$ | $D^v$ | $\mu_g$ | $L$ |
| --- | --- | --- | --- | --- | --- | --- | --- | --- |
| 12.00 | 0.6930 | 6.000 | 2.079 | 10.00 | 1.000 | 0.0500 | 100.0 | 38.00 |

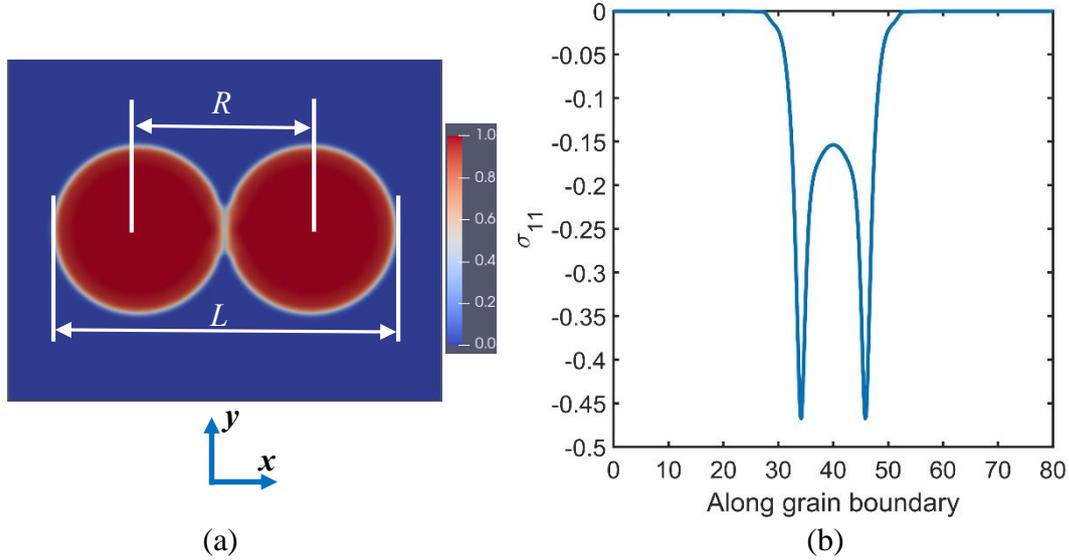

(a)  (b)

Figure 2: (a)The morphology of the two-particle model represented by $(C\eta_1)^2 + (C\eta_2)^2$ at 1200 timesteps and (b) the corresponding dimensionless normal stress ($\sigma_{11}$) distribution along the vertical grain boundary.

To our knowledge, PFMMS stands as the innovative phase-field-based model capable of capturing the intrinsic stress information along the grain boundary in solid-state sintering. Notably, PFMMS extends beyond just the grain boundary to capture stress distribution across the entire powdered compact, as shown in Fig. 3. The stress concentration at the neck region is captured. This holistic approach offers a significant advantage, as it enables the direct derivation of macroscale stress-strain rate information through micromechanics-based homogenization.

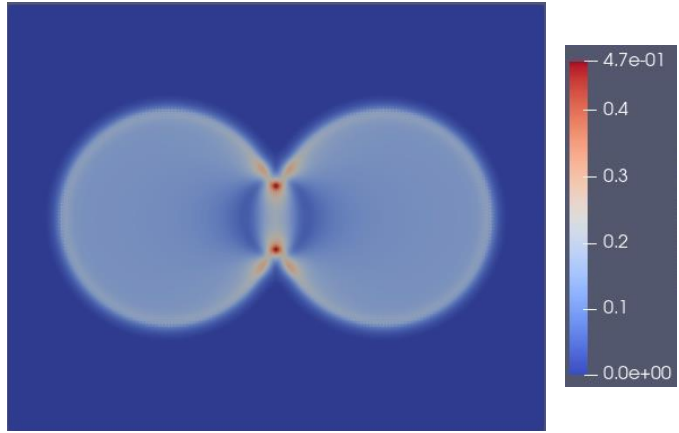

Figure 3: the distribution of Frobenius norm of the dimensionless stress tensor across the two-particle compact at 1200 timesteps.

### 3.1.2 Shrinkage dependence on $\frac{D^{gb}}{D^{sf}}$

Let $r_0$ denote the initial particle radius, $L$ represent the distance between the ends of the two-particle model, and $R$ be the distance between the particle centers, as illustrated in Fig. 2a. Let $\Delta L$ and $\Delta R$ denote the changes in $L$ and $R$, respectively, and $u$ represent the shrinkage of the particle centers due solely to grain motion. The detailed calculations of $\Delta L$, $\Delta R$, and $u$ in PFMMS are provided in Appendix B. Now, let $\Delta L^*$, $\Delta R^*$ and $u^*$ be the particle-radius-normalized counterparts of $\Delta L$, $\Delta R$, and $u$, respectively. Figure 4 plots $\Delta L^*$, $\Delta R^*$ and $u^*$ as functions of normalized time $t^*$ under different $\frac{D^{gb}}{D^{sf}}$ ratios when $\frac{\gamma^{gb}}{\gamma^{sf}} = 0$, as employed in [12]. The normalized time is defined as $t^* = \frac{\gamma^{sf} t}{\mu_g r_0}$. The normalized parameters utilized in this test are presented in Table 2.

Table 2 The normalized parameters employed in shrinkage simulations in PFMMS. $\mu_g$ is not an independent parameter and is inverse propositional to $D^{gb}$.

| Parameters $\frac{D^{gb}}{D^{sf}}$ | $\alpha$ | $\beta$ | $\kappa_C$ | $\kappa_\eta$ | $D^{sf}$ | $D^{gb}$ | $D^v$ | $\mu_g$ | $L$ |
|---|---|---|---|---|---|---|---|---|---|
| 1 | 12.00 | 0.00 | 6.00 | 0.00 | 100.00 | 100.00 | 0.05 | 200.0 | 0.00 |
| 0.1 | 12.00 | 0.00 | 6.00 | 0.00 | 100.00 | 10.00 | 0.05 | 2000.0 | 0.00 |
| 0.01 | 12.00 | 0.00 | 6.00 | 0.00 | 100.00 | 1.00 | 0.05 | 20000.0 | 0.00 |

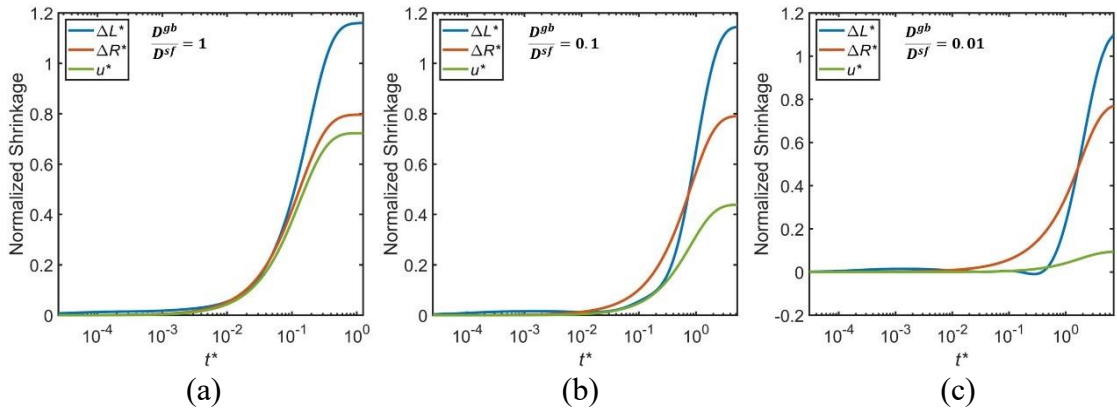

Figure 4: The normalized shrinkage as a function of dimensionless time for different $\frac{D^{gb}}{D^{sf}}$ ratios; $\Delta L^*$ represents the normalized distance change between the ends of the two particles, $\Delta R^*$ denotes the normalized distance change between the two particle centers, and $u^*$ stands for the shrinkage contribution induced by grain motion. The shrinkage is plotted on a semilogarithmic scale for better visibility of the differences.

The influence of the ratio $\frac{D^{gb}}{D^{sf}}$ on the contribution of $u^*$ to $\Delta R^*$ is notably significant. It's important to note that the final shrinkage $\Delta R^*$ at (meta-) equilibrium remains independent of $\frac{D^{gb}}{D^{sf}}$. However, the shrinkage contribution $u^*$ from grain motion is heavily reliant on $\frac{D^{gb}}{D^{sf}}$. Specifically, the shrinkage contribution from $u^*$ increases as $\frac{D^{gb}}{D^{sf}}$ goes up. It's noteworthy that the effect of $\frac{D^{gb}}{D^{sf}}$ on the shrinkage contribution $u^*$ due to grain motion in PFMMS aligns with observations from theoretical analyses in [4, 12]. Physically, a smaller $D^{gb}$ implies a larger effective viscosity $\mu_g$ for the grain phase, resulting in reduced grain motion and consequently less shrinkage. In extreme scenarios where $D^{gb}$ approaches zero, $u^*$ tends towards zero, as evident in Fig. 4c. However, even in such extreme cases, the shrinkage denoted by $\Delta L^*$ can still be achieved due to the contribution from pure surface diffusion-induced spherization.

In [12, 14, 15, 65, 66], it was noted that the shrinkage represented by $\Delta L^*$ may transiently decrease for specific $\frac{D^{gb}}{D^{sf}}$ ratios when the particle pair exhibits an elongated shape. This unusual phenomenon is attributed to surface diffusion and is similarly observed in PFMMS, as demonstrated in Fig. 4c. Furthermore, Fig. 5a displays the morphology of the two-particle model at the point where this anomalous behavior is observed, confirming the presence of an elongated shape. Additionally, Figure 5b shows the dimensionless energy $\mathcal{F}^*$ as a function of normalized time $t^*$. It is evident that $\mathcal{F}^*$ decreases monotonically as the microstructure evolves, thereby numerically confirming the proposed energy law defined in Eq. (13).

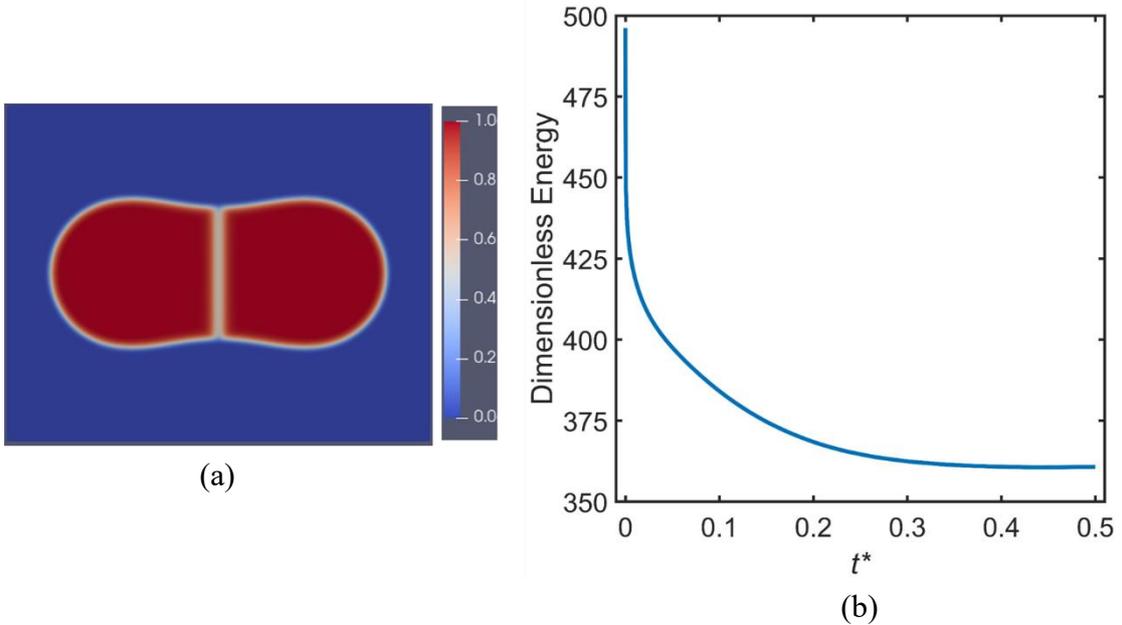

(a)                (b)

Figure 5: (a) The elongated shape when the shrinkage represented by $\Delta L^*$ decreases in Fig. 4c. The morphology is represented by $(C\eta_1)^2 + (C\eta_2)^2$; (b) the dimensionless energy as a function of dimensionless time $t^*$ during the simulation.

3.1.3 Shrinkage dependence on $\frac{\gamma^{gb}}{\gamma^{sf}}$

Figure 6 illustrates the evolution of $\Delta R^*$ and $u^*$ as functions of time under different ratios of $\frac{\gamma^{gb}}{\gamma^{sf}}$. The parameters employed in PFMMS for each energy ratio are detailed in Table 3. It is evident that the shrinkage $\Delta R^*$ is strongly influenced by $\frac{\gamma^{gb}}{\gamma^{sf}}$ and decreases as $\frac{\gamma^{gb}}{\gamma^{sf}}$ increases. This dependency arises because the energy ratio $\frac{\gamma^{gb}}{\gamma^{sf}}$ determines the final (meta-)equilibrium shape of the equally sized two-particle model, consequently impacting the final shrinkage.

In contrast, $u^*$ exhibits a relatively weaker dependence on $\frac{\gamma^{gb}}{\gamma^{sf}}$. This may be attributed to the fact that the shrinkage due to grain motion primarily occurs during the early stages of sintering. As $\frac{\gamma^{gb}}{\gamma^{sf}}$ mainly affects the final equilibrium shape during the later stages of sintering, when diffusion dominates over grain motion, it has a greater impact on the shrinkage contribution due to diffusion rather than grain motion.

Table 3 The normalized parameters employed in simulations of the energy ratio effect in PFMMS.

| $\frac{\gamma^{gb}}{\gamma^{sf}}$ \ Parameters | $\alpha$ | $\beta$ | $\kappa_C$ | $\kappa_\eta$ | $D^{sf}$ | $D^{gb}$ | $D^v$ | $\mu_g$ | $L$ |
|---|---|---|---|---|---|---|---|---|---|
| 0 | 12.00 | 0.00 | 6.00 | 0.00 | 100.00 | 10.00 | 0.05 | 5000.0 | 0.00 |
| 0.51 | 12.00 | 0.51 | 6.00 | 1.53 | 100.00 | 10.00 | 0.05 | 5000.0 | 38.00 |
| 0.80 | 12.00 | 0.80 | 6.00 | 2.40 | 100.00 | 10.00 | 0.05 | 5000.0 | 38.00 |
| 1.0 | 12.00 | 1.00 | 6.00 | 3.00 | 100.00 | 10.00 | 0.05 | 5000.0 | 38.00 |

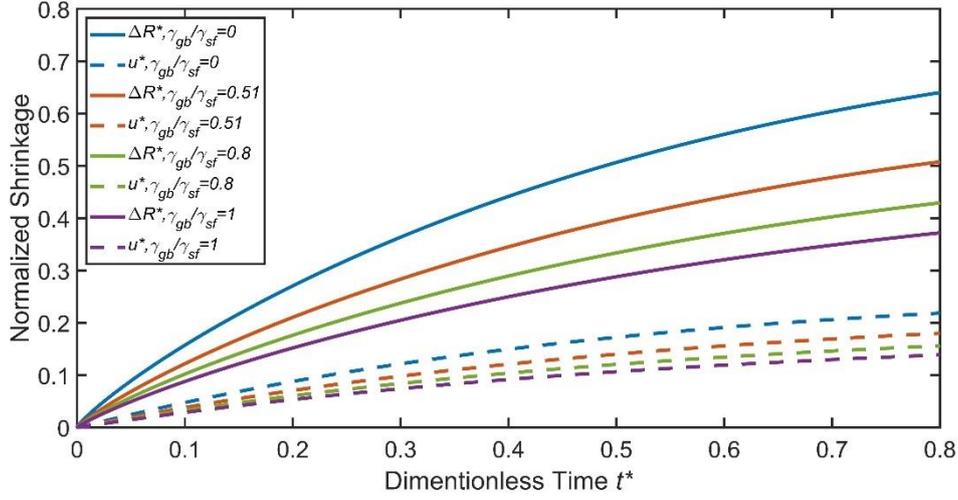

Figure6: The normalized shrinkages represented by respective $\Delta R^*$ and $u^*$ as functions of the normalized time $t^*$ under different $\frac{\gamma^{gb}}{\gamma^{sf}}$ ratios. $\Delta R^*$ denotes the normalized distance change between the two particle centers, and $u^*$ stands for the normalized shrinkage contribution induced by grain motion.

3.1.4 Neck growth and velocity distribution

In Fig. 7, we present the evolution of the normalized neck radius as a function of normalized time, using the same tests outlined in Section 3.1.3. The neck growth, assuming a rigid body, can be approximated by the equation:

$$\frac{X}{r_0} = Bt^n \qquad (34)$$

Here, $X$ represents the neck radius, $n$ and $B$ are numerical constants. The detailed calculation of $X$ is provided in Appendix B. Table 4 displays the exponent $n$ and the goodness of fit $R^2$ for each $\frac{\gamma^{gb}}{\gamma^{sf}}$ ratio.

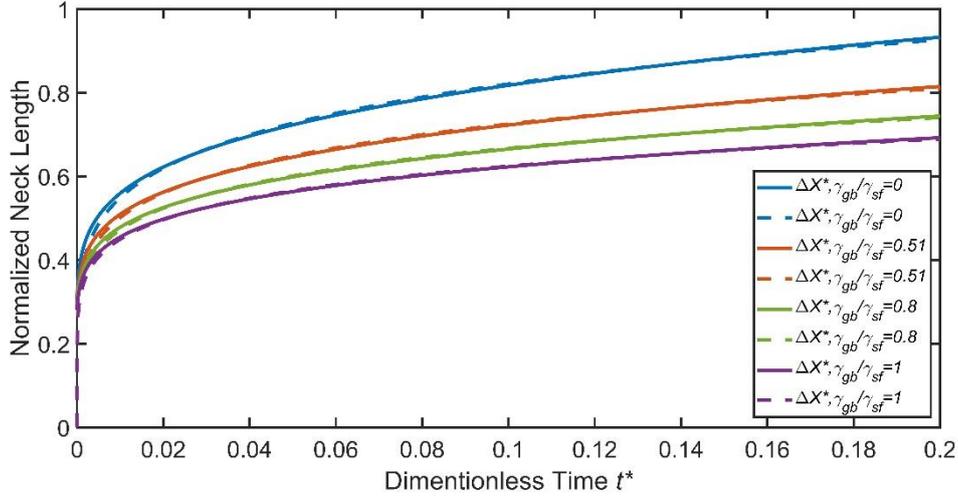

Figure 7: the evolution of the normalized neck radius as a function of different $\frac{\gamma^{gb}}{\gamma^{sf}}$ ratios; the solid lines represent results from the simulations and the dashed lines are from the corresponding fitting.

Table 4 The exponent *n* and the goodness of fit $R^2$ for each energy ratio

| $\frac{\gamma^{gb}}{\gamma^{sf}}$ | $n$ | $R^2$ |
|---|---|---|
| 0 | 0.1749 | 0.9983 |
| 0.51 | 0.1602 | 0.9986 |
| 0.80 | 0.1504 | 0.9988 |
| 1.0 | 0.1427 | 0.9989 |

The theoretical analysis, assuming a rigid body, yields a value of approximately 0.1667 for the parameter *n*. The discrepancy in *n* values between the rigid-body analysis and the employed PFMMS may stem from their differing assumptions regarding grain motion: one assumes grains to be rigid, while the other considers them deformable. Figure 8a shows the dimensionless *x*-velocity distribution along the horizontal direction, represented as $\dot{u}_1$, across the compact at 3000 timesteps when $\frac{\gamma^{gb}}{\gamma^{sf}} = 0.51$. To enhance visualization of the velocity profile within the particle compact, the velocity distribution is obtained by calculating the product $C\dot{u}_1$. Figure 8b displays the distribution of $\dot{u}_1$ along the center horizontal line (the dashed white line). It is observed that, far from the grain boundary, $\dot{u}_1$ is nearly uniform, exhibiting quasi-rigid body motion. In contrast, near the grain boundary, $\dot{u}_1$ demonstrates a graded behavior. It is noteworthy that previous theoretical work [15] demonstrated that the exponent $n$ is inversely proportional to $\frac{\gamma^{gb}}{\gamma^{sf}}$. This relationship is also qualitatively evident in our PFMMS results (see Table 4), where $n$ decreases as $\frac{\gamma^{gb}}{\gamma^{sf}}$ increases. This trend arises because higher $\gamma^{gb}$ values represent greater grain boundary tension, thereby impeding sintering.

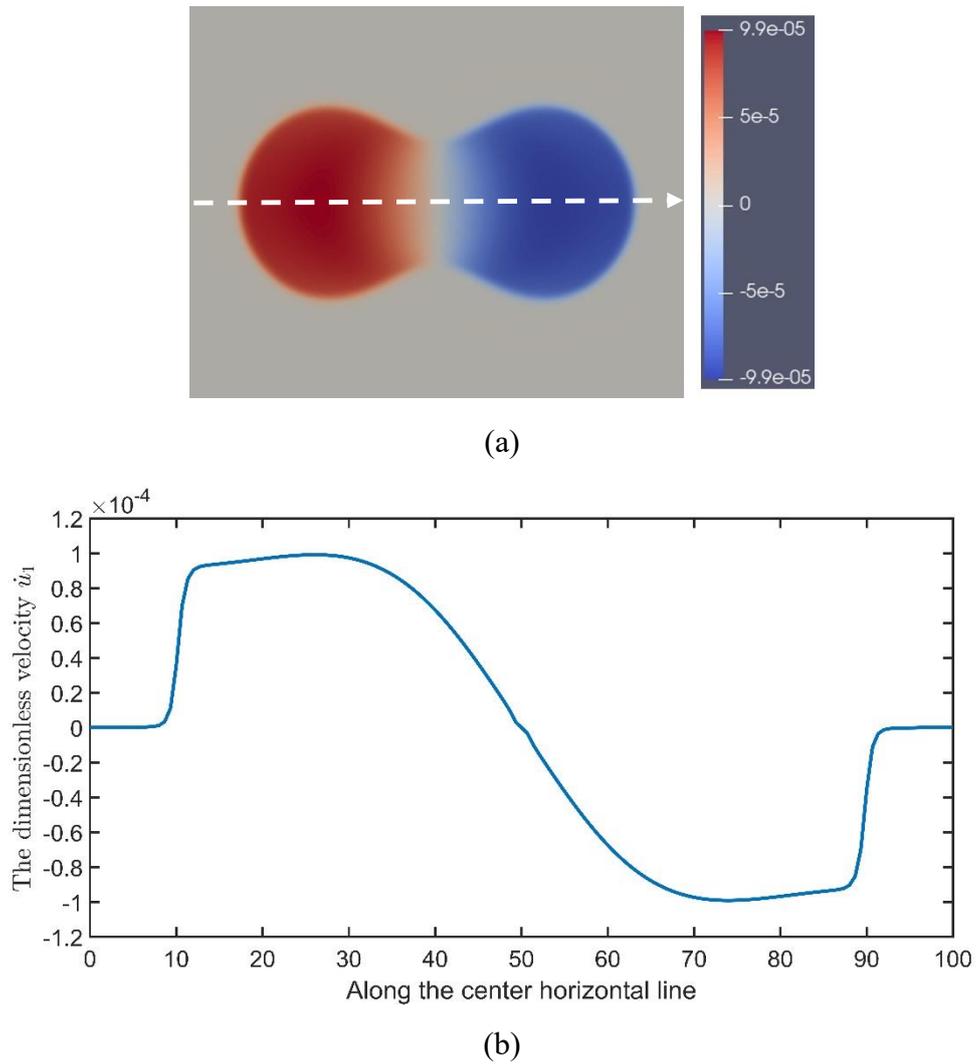

(a)

(b)

Figure 8: (a) the distribution of the dimensionless horizontal velocity component $\dot{u}_1$ across the powder compact and (b) the distribution of $\dot{u}_1$ along the center horizontal line (the dashed white line) at 3000 timesteps.

Further enhancements within the PFMMS framework could bring its predictions closer to those of rigid-body analysis. For instance, incorporating non-uniform grain phase viscosity $\mu_g$ within grain and grain boundaries may prove beneficial, a direction we intend to explore in future studies.

3.2 Particle chain model

In a recent study [5], a novel particle-chain benchmark test was introduced to assess the accuracy of a sintering model. This test evaluates whether shrinkage remains consistent regardless of system size. The rationale behind this approach is as follows: assuming each grain boundary operates independently, the rate of densification at a given time should not vary with the number of grain boundaries in the chain. In other words, the densification strain-time curves should converge into a single master curve as the number of grains reaches a certain threshold. To further validate the effectiveness of the proposed PFMMS, we utilize the particle chain model described in [5], which comprises varying numbers of equally sized grains. The employed parameters in this study are given in Table 5.

Table 5 The normalized parameters employed in chain model simulations.

| $\alpha$ | $\beta$ | $\kappa_C$ | $\kappa_\eta$ | $D^{sf}$ | $D^{gb}$ | $D^v$ | $\mu_g$ | $L$ |
|---|---|---|---|---|---|---|---|---|
| 12.00 | 0.5100 | 6.000 | 1.530 | 100.0 | 100.0 | 0.5000 | 200.0 | 38.00 |

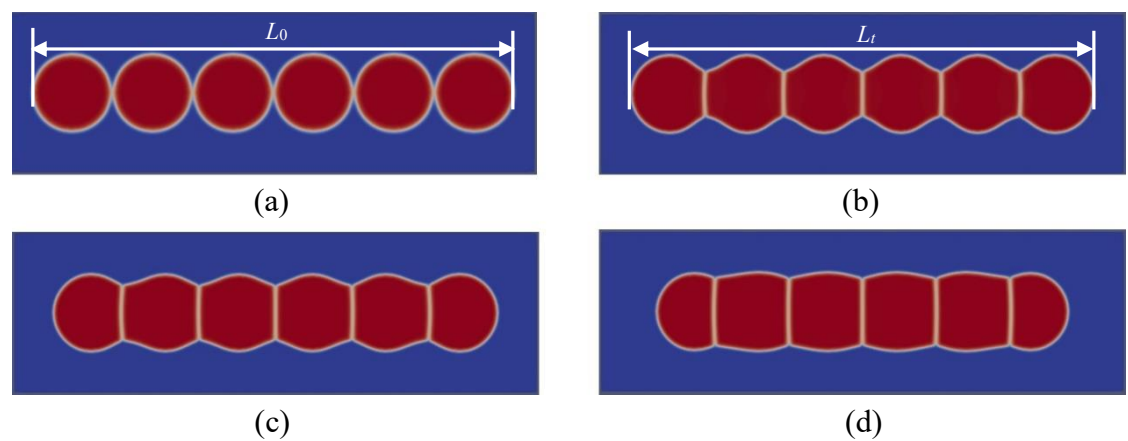

(a)　　　　　　　　　　　　　　　　(b)

(c)　　　　　　　　　　　　　　　　(d)

Figure9: The microstructure morphology of the 6-grain chain at (a) initial, (b)1000, (c) 2000 and (d) 4000 time steps. $L_0$ and $L_t$ represent the respective initial and current distances between the two ends of the particle chain.

In Fig. 9, we use the 6-grain chain model as an illustration. Let $L(t_0)$ and $L(t)$ represent the initial and current distances, respectively, between the ends of the leftmost and rightmost grains along the horizontal $x$ axis. The absolute densification strain is defined as:

$$\varepsilon_x(t) = \frac{L(t_0)-L(t)}{L(t_0)} \tag{35}$$

Figure 9 depicts the microstructure evolution of the 6-grain model, showcasing both neck growth and shrinkage. Additionally, we employ three other chain models with 2, 4, and 8 grains, respectively, and display the densification strain-time curves for each model in Fig. 10. Notably, these curves converge as the number of grains within the chain increases. This convergence suggests that the densification strain remains consistent regardless of system size, thereby validating the efficacy of the proposed PFMMS.

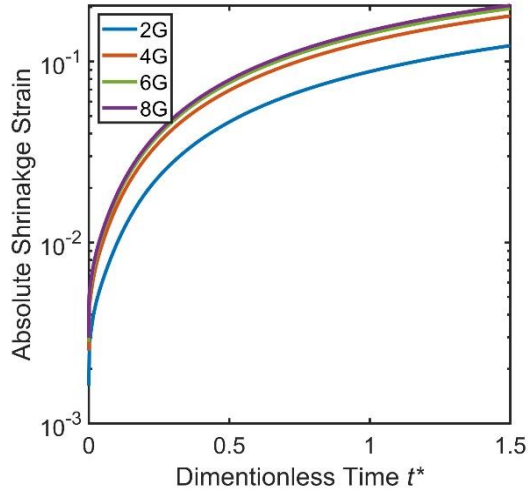

Figure 10: The densification strain $\varepsilon_x$ as a function of the normalize time $t^*$ for the respective 2-grain (2G), 4-grain(4G), 6-grain(6G) and 8-grain chains(8G).

## 4. Discussions

The numerical examples presented in Section 3 demonstrate the effectiveness of the proposed PFMMS in capturing key aspects of solid-state sintering, such as stress distribution, and shrinkage. In this section, we delve into the inherent features of PFMMS that facilitate these capabilities.

(1) Mechanics of grain motion

PFMMS's ability to capture stress information is encapsulated in Eq. (24), which governs grain motion. Notably, our simulations operate under free sintering conditions without external stress. Internal stress in free sintering arises from intrinsic surface tension and grain boundary tension, as indicated in Eq. (27). In cases where external stress is applied, such as in pressure-assisted sintering, Eq. (24) remains valid, with the external stress effect being incorporated through boundary conditions. Consequently, the governing equations in PFMMS apply to both free sintering and pressure-assisted sintering.

Furthermore, the constitutive model governing the mechanical behavior of grains is not confined to the forms presented in Eqs. (19) and (26) in this study. For instance, the inclusion of energy dissipation $\mathcal{D}_m$ through dislocation motion can be utilized to describe the mechanical behavior of grains in metal powder sintering. Thus, PFMMS can be adapted to accommodate various mechanical constitutive models.

(2) Sintering Dependence on $\frac{D^{gb}}{D^{sf}}$ and $\frac{\gamma^{gb}}{\gamma^{sf}}$

The incorporation of $\frac{D^{gb}}{D^{sf}}$ dependence is facilitated by the inverse proportionality between grain phase viscosity $\mu_g$ and grain boundary diffusion $D^{gb}$. A higher $D^{gb}$ leads to lower viscosity, reducing resistance to mass transport via grain motion and increasing the contribution of convection flux in Eq. (22).

Similarly, the $\frac{\gamma^{gb}}{\gamma^{sf}}$ dependence is embedded in the intrinsic body force **b** in Eq. (27), which accounts for contributions from both surface tension and grain boundary tension. A greater $\gamma^{gb}$ results in increased grain boundary tension, resisting grain motion through mechanical equilibrium and thus reducing shrinkage.

(3) Thermodynamic Equilibrium State:

In [5], it is emphasized that besides the condition of shrinkage strain independence from system size, as depicted in Fig. 10, another critical criterion for a physically consistent sintering model is the preservation of the thermodynamic equilibrium state. The energy law (Eq. (13)) in PFMMS demonstrates that while the dissipation $\mathcal{D}_m$ from grain motion mechanics accelerates sintering by introducing convection flux in Eq. (22), it does not alter the thermodynamic equilibrium of the process.

Furthermore, considering the non-negativity of various energy dissipation terms in Eqs. (15), (16), and (19), the free energy of the sintering system is guaranteed to decrease as the microstructure evolves. Consequently, phenomena such as non-densification, which increases the system's free energy, as reported in [5], are precluded.

## 5. Conclusion

In this study, we introduce a novel phase-field-micromechanics model of sintering (PFMMS) that addresses a longstanding challenge in phase-field modeling: incorporating grain motion's contribution to densification in a thermodynamically consistent manner. The key innovation of PFMMS is its unified energy law, which is utilized to rigorously derive governing equations via variational principles. This approach establishes the thermodynamic driving force for grain motion and its impact on densification from the system's free energy. Consequently, PFMMS ensures microstructure evolution in a direction that minimizes system energy, eliminating non-densifying phenomena as observed and documented in [5].

Validation of the PFMMS framework utilizes benchmark examples from recent theoretical and numerical analyses, incorporating the interfacially consistent phase-field free energy functional proposed in contemporary literature. PFMMS captures stress distribution along and beyond grain boundaries, exhibits system-size-independent shrinkage strain, and maintains thermodynamic equilibrium states. These features are fundamental requirements for a physically consistent sintering model.

Moving forward, we aim to extend the PFMMS to pressure-assisted sintering and multiphase material sintering, incorporating grain boundary energy anisotropy to study grain rotation and expanding into multi-physics sintering scenarios. Additionally, we foresee potential applications of PFMMS in developing thermodynamically consistent phase-field models for material processes where mechanics significantly contribute to the overall behavior.

**Acknowledgment**

Q.Y. acknowledges the financial support from the Science and Technology Commission of Shanghai Municipality through grant No. 23010500400, the key program (No. 52232002), and the grant (No. 12272214) from the Natural Science Foundation of China.

**Appendix A**

As defined in Section 2, the free energy functional $\mathcal{F}$ and the energy dissipation rate $\mathcal{D}$ of the powdered system may generally be expressed in terms of the introduced phase-fields $C$ and $\eta_i$ as follows:

$$\mathcal{F} = \int_\Omega f\left(C, \eta_1, \eta_1, \ldots, \eta_{n_g}, \nabla C, \nabla \eta_1, \nabla \eta_2, \ldots, \nabla \eta_{n_g}\right) d\Omega \tag{A1}$$

$$\mathcal{D} = \mathcal{D}_C + \sum_i^{n_g} \mathcal{D}_{\eta_i} + \mathcal{D}_m \tag{A2}$$

where $n_g$ denotes the total number of grains with different orientations, and $\mathcal{D}_C$, $\mathcal{D}_{\eta_i}$ and $\mathcal{D}_m$ represent the energy dissipation mechanisms due to the respective diffusions, grain boundary evolution, and grain motion. They are defined as follows:

$$\mathcal{D}_C = \int_\Omega \frac{1}{2} C^2 M^{-1} \|\dot{\mathbf{u}}_g^*\|^2 d\Omega \tag{A3}$$

$$\mathcal{D}_{\eta_i} = \int_\Omega \frac{1}{2} L^{-1} \|R_{\eta_i}^t\|^2 d\Omega \tag{A4}$$

$$\mathcal{D}_m = \int_\Omega \frac{1}{2} \mu^{eff} \|\nabla \dot{\mathbf{u}} + (\nabla \dot{\mathbf{u}})^{\mathrm{T}}\|^2 d\Omega \tag{A5}$$

where $R_{\eta_i}^t$ represents the rate of change of $\eta_i$ along the characteristic trajectory defined by mechanics velocity $\dot{\mathbf{u}}$, i.e. the materials derivative of $\eta_i$ in continuum mechanics, and is given as:

$$R_{\eta_i}^t = \frac{\mathrm{D}\eta_i}{\mathrm{D}t} = \frac{\partial \eta_i}{\partial t} + \dot{\mathbf{u}} \cdot \nabla \eta_i \tag{A6}$$

Additionally, the kinematic constraints, i.e., Eqs. (11) and (8) in section 2.1 governing local behavior of phase field $C$ and mechanics velocity $\dot{\mathbf{u}}$ must be enforced:

$$\frac{\partial C}{\partial t} + \nabla \cdot (C\dot{\mathbf{u}}) + \nabla \cdot (C\dot{\mathbf{u}}_g^*) = 0 \tag{A7}$$

$$\nabla \cdot \dot{\mathbf{u}} = \mathbf{0} \tag{A8}$$

To perform variation, the flow maps $\mathbf{x}(\boldsymbol{\xi}, t)$ corresponding to the mechanics velocity $\dot{\mathbf{u}}$, $\mathbf{x}_C(\boldsymbol{\xi}, t)$ corresponding to the drift velocity $\dot{\mathbf{u}}_g^*$, and $R_{\eta_i}(t)$ consistent with the gradient flow of $\eta_i$ are defined as follows:

$$\mathbf{x}(\boldsymbol{\xi}, t): \frac{\partial \mathbf{x}}{\partial t}(\boldsymbol{\xi}, t) = \dot{\mathbf{u}}(\mathbf{x}(\boldsymbol{\xi}, t), t) \tag{A9}$$

$$\mathbf{x}_C(\boldsymbol{\xi}, t): \frac{\partial \mathbf{x}_C}{\partial t}(\boldsymbol{\xi}, t) = \dot{\mathbf{u}}_g^*(\mathbf{x}_C(\boldsymbol{\xi}, t), t) \tag{A10}$$

$$R_{\eta_i}(t): \frac{\partial R_{\eta_i}}{\partial t} = R_{\eta_i}^t \tag{A11}$$

where $\xi$ represents a material point in Lagrangian description. In the following, we will show that the balance of mechanics equation, the extended Cahn-Hillard equation, and the Allen-Cahn equation can be derived by performing variation of the defined free energy functional $\mathcal{F}$ with respect to the introduced flow maps $x$, $x_C$, and $R_{\eta_i}$.

The principle of virtual work [67] allows us to avoid performing multiple variations and having to change variables between Lagrangian and Eulerian descriptions, and results in one large variational calculation for each energy component. To employ the principle of virtual work, we need to translate the kinematic constraints defined by Eqs. (A6)-(A8) into variational relations as follows:

$$\delta C + \nabla \cdot (C \delta x) + \nabla \cdot (C \delta x_C) = 0 \tag{A12}$$

$$\delta \eta_i + \delta x \cdot \nabla \eta_i - \delta R_{\eta_i} = 0 \tag{A13}$$

$$\nabla \cdot (\delta x) = 0 \tag{A14}$$

Here $\delta x, \delta x_C,$ and $\delta R_{\eta_i}$ are infinitesimal increments defining the direction of variation. Then the variation of the defined free energy functional $\mathcal{F}$ can be expressed as follows:

$$\delta \mathcal{F} = \int_\Omega \left[ \frac{\partial f}{\partial C} \delta C + \frac{\partial f}{\partial \nabla C} \cdot \nabla \delta C + \sum_i \frac{\partial f}{\partial \eta_i} \delta \eta_i + \frac{\partial f}{\partial \nabla \eta_i} \cdot \nabla \delta \eta_i \right] d\Omega \tag{A15}$$

According to Eqs. (A12) and (A13), we have:

$$\delta \mathcal{F} = \int_\Omega \left[ \frac{\partial f}{\partial C} \left( -\nabla \cdot (C \delta x) - \nabla \cdot (C \delta x_C) \right) + \frac{\partial f}{\partial \nabla C} \cdot \nabla \left( -\nabla \cdot (C \delta x) - \nabla \cdot (C \delta x_C) \right) + \right.$$
$$\left. \sum_i \frac{\partial f}{\partial \eta_i} \left( -\delta x \cdot \nabla \eta_i + \delta R_{\eta_i} \right) + \frac{\partial f}{\partial \nabla \eta_i} \cdot \nabla \left( -\delta x \cdot \nabla \eta_i + \delta R_{\eta_i} \right) \right] d\Omega \tag{A16}$$

Applying chain rule and divergence theorem and imposing Eq. (14), we have:

$$\delta \mathcal{F} = -\int_\Omega \nabla f \cdot \delta x \ d\Omega - \int_\Omega \left( \frac{\partial f}{\partial \nabla C} \cdot \nabla \delta x \cdot \nabla C + \sum_i \frac{\partial f}{\partial \nabla \eta_i} \cdot \nabla \delta x \cdot \nabla \eta_i \right) d\Omega -$$
$$\int_\Omega \frac{\delta \mathcal{F}}{\delta C} \nabla \cdot (C \delta x_C) \ d\Omega + \int_\Omega \sum_i \frac{\delta \mathcal{F}}{\delta \eta_i} \delta R_{\eta_i} \ d\Omega \tag{A17}$$

In deriving Eq. (A17), the boundary integrals from the divergence theorem are discarded by imposing appropriate boundary conditions, which is consistently employed in the derivation. Additionally, the following identities are used:

$$\nabla f = \frac{\partial f}{\partial C}\nabla C + \left(\frac{\partial f}{\partial \nabla C}\cdot \nabla\right)\nabla C + \Sigma_i \frac{\partial f}{\partial \eta_i}\nabla \eta_i + \left(\frac{\partial f}{\partial \nabla \eta_i}\cdot \nabla\right)\nabla \eta_i \quad (A18)$$

$$\frac{\delta \mathcal{F}}{\delta C} = \frac{\partial f}{\partial C} - \nabla \cdot \frac{\partial f}{\partial \nabla C} \quad (A19)$$

$$\frac{\delta \mathcal{F}}{\delta \eta_i} = \frac{\partial f}{\partial \eta_i} - \nabla \cdot \frac{\partial f}{\partial \nabla \eta_i} \quad (A20)$$

Further applying the divergence theorem, Eq. (A17) can be simplified as:

$$\delta \mathcal{F} = \int_\Omega \left(-\nabla f + \nabla \cdot \left(\frac{\partial f}{\partial \nabla C}\otimes \nabla C + \Sigma_i \frac{\partial f}{\partial \nabla \eta_i}\otimes \nabla \eta_i\right)\right)\cdot \delta x\, d\Omega + \int_\Omega C\nabla \frac{\delta \mathcal{F}}{\delta C}\cdot \delta \mathbf{x}_C\, d\Omega +$$

$$\int_\Omega \Sigma_i \frac{\delta \mathcal{F}}{\delta \eta_i}\delta R_{\eta_i}\, d\Omega \quad d\Omega \quad (A21)$$

This results in the following conservative forces:

$$\frac{\delta \mathcal{F}}{\delta x} = -\nabla f + \nabla \cdot \left(\frac{\partial f}{\partial \nabla C}\otimes \nabla C + \Sigma_i \frac{\partial f}{\partial \nabla \eta_i}\otimes \nabla \eta_i\right) \quad (A22)$$

$$\frac{\delta \mathcal{F}}{\delta \mathbf{x}_C} = C\nabla \frac{\delta \mathcal{F}}{\delta C} \quad (A23)$$

$$\frac{\delta \mathcal{F}}{\delta R_{\eta_i}} = \frac{\delta \mathcal{F}}{\delta \eta_i} \quad (A24)$$

Next, the variation of the energy dissipation rate $\mathcal{D}$ with respect to the mechanics velocity $\dot{\mathbf{u}}$, the drift velocity $\dot{\mathbf{u}}_g^*$, and $R_{\eta_i}^t$ generates the respective dissipative forces:

$$\frac{\delta \mathcal{D}}{\delta \dot{\mathbf{u}}} = -\nabla \cdot \left(\mu^{eff}(\nabla \dot{\mathbf{u}} + (\nabla \dot{\mathbf{u}})^T)\right) \quad (A25)$$

$$\frac{\delta \mathcal{D}}{\delta \dot{\mathbf{u}}_g^*} = C^2 M^{-1}\dot{\mathbf{u}}_g^* \quad (A26)$$

$$\frac{\delta \mathcal{D}}{\delta R_{\eta_i}^t} = L^{-1}R_{\eta_i}^t \quad (A27)$$

Additionally, the Lagrange multiplier is needed to account for the constraint $\nabla \cdot \dot{\mathbf{u}} = 0$, which generates the following variation:

$$\delta \int_\Omega p\, \nabla \cdot \dot{\mathbf{u}}\, d\Omega = \int_\Omega (-\nabla p)\cdot \delta \dot{\mathbf{u}}\, d\Omega \quad (A28)$$

where the divergence theorem is applied. Now, assembling force balance equations in the form:

$$force_{conservative} + force_{dissipative} = 0 \quad (A29)$$

We arrive at the following relations:

$$-\left[\nabla \cdot \left(\mu^{eff}(\nabla \dot{\mathbf{u}} + (\nabla \dot{\mathbf{u}})^T)\right) + \nabla p\right] - \nabla f + \nabla \cdot \left(\frac{\partial f}{\partial \nabla C}\otimes \nabla C + \Sigma_i \frac{\partial f}{\partial \nabla \eta_i}\otimes \nabla \eta_i\right) = 0 \quad (A30)$$

$$C\dot{\mathbf{u}}_g^* = -M\nabla \frac{\delta \mathcal{F}}{\delta C} \quad (A31)$$

$$R^t_{\eta_i} = -L\frac{\delta \mathcal{F}}{\delta \eta_i} \tag{A32}$$

Note that Eq. (A30) additionally contains the Lagrange multiplier term $p$ which effectively projects the forces to not violate the constraint defined in Eq. (A8). Finally, substituting $C\dot{\mathbf{u}}^*_g$ and $R^t_{\eta_i}$ back into Eqs. (A6) and (A7), the governing equations in the proposed PFMMS framework are summarized as follows:

$$\frac{\partial C}{\partial t} + \nabla \cdot (C\dot{\mathbf{u}}) = \nabla \cdot \left(M\nabla \frac{\delta \mathcal{F}}{\delta C}\right) \tag{A33}$$

$$\frac{\partial \eta_i}{\partial t} + \dot{\mathbf{u}} \cdot \nabla \eta_i = -L\frac{\delta \mathcal{F}}{\delta \eta_i} \tag{A34}$$

$$\nabla \cdot \boldsymbol{\sigma} + \mathbf{b} = \mathbf{0} \tag{A35}$$

$$\nabla \cdot \dot{\mathbf{u}} = \mathbf{0} \tag{A36}$$

where the Cauchy stress tensor $\boldsymbol{\sigma}$ and the body force $\mathbf{b}$ are defined as follows:

$$\boldsymbol{\sigma} = \mu^{eff}(\nabla \dot{\mathbf{u}} + (\nabla \dot{\mathbf{u}})^{\mathrm{T}}) + p\mathbf{I} \tag{A37}$$

$$\mathbf{b} = \nabla f - \nabla \cdot \left(\frac{\partial f}{\partial \nabla C} \otimes \nabla C + \sum_i \frac{\partial f}{\partial \nabla \eta_i} \otimes \nabla \eta_i\right) \tag{A38}$$

**Appendix B**

For the employed two-particle model shown in Fig. B1, let $R$ denote the $x$-distance between particle centers, $L$ represent the $x$-distance between the ends, $u$ be the shrinkage of the particle centers due solely to grain motion, and $X$ denote the neck radius.

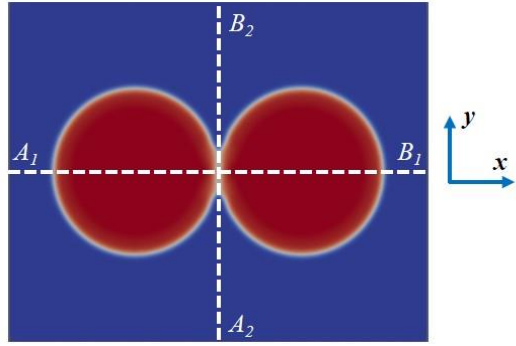

Figure B1: the schematic illustration of the employed two-particle model

The $x$-coordinate of the left particle center (represented as $x_{Center}^1$) and the $x$-coordinate of the right particle center (denoted as $x_{Center}^2$) are calculated using the following formulas:

$$x_{Center}^1 = \frac{\int_\Omega C\eta_1 x \, d\Omega}{\int_\Omega C\eta_1 \, d\Omega} \tag{B1}$$

$$x_{Center}^2 = \frac{\int_\Omega C\eta_2 x \, d\Omega}{\int_\Omega C\eta_2 \, d\Omega} \tag{B2}$$

Then the distance $R$ and its change $\Delta R$ at a given time $t$ are evaluated as follows:

$$R(0) = x_{Center}^2(0) - x_{Center}^1(0) \tag{B3}$$

$$R(t) = x_{Center}^2(t) - x_{Center}^1(t) \tag{B4}$$

$$\Delta R = R(0) - R(t) \tag{B5}$$

At a given time $t$, the length $L$ is calculated by integrating $C$ over the horizontal line segment $\overline{A_1 B_1}$ in the middle of the domain:

$$L(t) = \int_{\overline{A_1 B_1}} C \, ds \tag{B6}$$

Then the shrinkage $\Delta L$ is calculated as follows:

$$\Delta L = L(0) - L(t) \tag{B7}$$

Similarly, to calculate the neck radius $X$ at a given time $t$, $C$ is integrated over the vertical line segment $\overline{A_2 B_2}$ in the middle of the domain and then halved:

$$X(t) = \frac{1}{2} \int_{\overline{A_2 B_2}} C \, ds \tag{B8}$$

Let $\overline{\dot{u}_x^1}$ and $\overline{\dot{u}_x^2}$ be the volume-averaged velocity components of the left and right particles, respectively, in the *x*-direction. They are evaluated as follows:

$$\overline{\dot{u}_x^1}(t) = \frac{\int_\Omega C\eta_1 \dot{u}_x \, d\Omega}{\int_\Omega C\eta_1 \, d\Omega} \tag{B9}$$

$$\overline{\dot{u}_x^2}(t) = \frac{\int_\Omega C\eta_2 \dot{u}_x \, d\Omega}{\int_\Omega C\eta_2 \, d\Omega} \tag{B10}$$

where $\dot{u}_x$ is the *x*-component of the mechanics velocity $\dot{\mathbf{u}}$. The relative *x*-velocity $\overline{\dot{u}_x}$ between the two particles is given as:

$$\overline{\dot{u}_x} = \overline{\dot{u}_x^1} - \overline{\dot{u}_x^2} \tag{B11}$$

Then at a given time *t*, the shrinkage contribution *u* due to grain motion is calculated by integrating the relative *x*-velocity over time:

$$u(t) = \int_0^t \overline{\dot{u}_x} \, d\tau \tag{B12}$$

References

[1] R. German, Sintering: from empirical observations to scientific principles, Butterworth-Heinemann, 2014.

[2] R.K. Bordia, S.J.L. Kang, E.A. Olevsky, Current understanding and future research directions at the onset of the next century of sintering science and technology, Journal of the American Ceramic Society, 100 (2017) 2314-2352.

[3] R. Grupp, M. Nöthe, B. Kieback, J. Banhart, Cooperative material transport during the early stage of sintering, Nature Communications, 2 (2011) 1-6.

[4] F. Wakai, G. Okuma, Rigid body motion of multiple particles in solid-state sintering, Acta Materialia, (2022) 118092.

[5] M. Seiz, Effect of rigid body motion in phase-field models of solid-state sintering, Computational Materials Science, 215 (2022) 111756.

[6] R. Shi, M. Wood, T.W. Heo, B.C. Wood, J. Ye, Towards understanding particle rigid-body motion during solid-state sintering, Journal of the European Ceramic Society, 41 (2021) 211-231.

[7] Y.U. Wang, Computer modeling and simulation of solid-state sintering: A phase field approach, Acta materialia, 54 (2006) 953-961.

[8] J. Frenkel, Viscous flow of crystalline bodies under the action of surface tension, J. Phys.(USS R), 9 (1945) 385.

[9] W.D. Kingery, M. Berg, Study of the initial stages of sintering solids by viscous flow, evaporation-condensation, and self-diffusion, Journal of Applied Physics, 26 (1955) 1205-1212.

[10] R.L. Coble, Sintering crystalline solids. I. Intermediate and final state diffusion models, Journal of applied physics, 32 (1961) 787-792.

[11] W. Coblenz, J. Dynys, R. Cannon, R. Coble, Initial stage solid state sintering models. A critical analysis and assessment, Sintering Processes. Materials Science Research, 13 (1980) 141-157.

[12] F. Wakai, K. Brakke, Mechanics of sintering for coupled grain boundary and surface diffusion, Acta materialia, 59 (2011) 5379-5387.

[13] F. Wakai, O. Guillon, G. Okuma, N. Nishiyama, Sintering forces acting among particles during sintering by grain-boundary/surface diffusion, Journal of the American Ceramic Society, 102 (2019) 538-547.

[14] W. Zhang, I. Gladwell, Sintering of two particles by surface and grain boundary diffusion–a three-dimensional model and a numerical study, Computational materials science, 12 (1998) 84-104.

[15] W. Zhang, J. Schneibel, The sintering of two particles by surface and grain boundary diffusion—a two-dimensional numerical study, Acta metallurgica et materialia, 43 (1995) 4377-4386.

[16] F. Parhami, R. McMeeking, A network model for initial stage sintering, Mechanics of materials, 27 (1998) 111-124.

[17] S. Luding, K. Manetsberger, J. Müllers, A discrete model for long time sintering, Journal of the Mechanics and Physics of Solids, 53 (2005) 455-491.

[18] C. Martin, L. Schneider, L. Olmos, D. Bouvard, Discrete element modeling of metallic powder sintering, Scripta materialia, 55 (2006) 425-428.

[19] G.N. Hassold, I.W. Chen, D.J. Srolovitz, Computer simulation of final-stage sintering: I, model kinetics, and microstructure, Journal of the American Ceramic Society, 73 (1990) 2857-2864.

[20] V. Tikare, M. Braginsky, E.A. Olevsky, Numerical simulation of solid-state sintering: I, sintering of three particles, Journal of the American Ceramic Society, 86 (2003) 49-53.

[21] M. Braginsky, V. Tikare, E. Olevsky, Numerical simulation of solid state sintering, International

journal of solids and structures, 42 (2005) 621-636.

[22] R. Bjørk, H.L. Frandsen, V. Tikare, E. Olevsky, N. Pryds, Strain in the mesoscale kinetic Monte Carlo model for sintering, Computational materials science, 82 (2014) 293-297.

[23] V. Kumar, Z. Fang, P. Fife, Phase field simulations of grain growth during sintering of two unequal-sized particles, Materials Science and Engineering: A, 528 (2010) 254-259.

[24] J. Deng, A phase field model of sintering with direction-dependent diffusion, Materials transactions, 53 (2012) 385-389.

[25] K. Shinagawa, S. Maki, K. Yokota, Phase-field simulation of platelike grain growth during sintering of alumina, Journal of the European Ceramic Society, 34 (2014) 3027-3036.

[26] S. Biswas, D. Schwen, J. Singh, V. Tomar, A study of the evolution of microstructure and consolidation kinetics during sintering using a phase field modeling based approach, Extreme Mechanics Letters, 7 (2016) 78-89.

[27] K. Chockalingam, V. Kouznetsova, O. Van der Sluis, M. Geers, 2D Phase field modeling of sintering of silver nanoparticles, Computer Methods in Applied Mechanics and Engineering, 312 (2016) 492-508.

[28] S. Biswas, D. Schwen, H. Wang, M. Okuniewski, V. Tomar, Phase field modeling of sintering: Role of grain orientation and anisotropic properties, Computational materials science, 148 (2018) 307-319.

[29] X. Zhang, Y. Liao, A phase-field model for solid-state selective laser sintering of metallic materials, Powder technology, 339 (2018) 677-685.

[30] F. Abdeljawad, D.S. Bolintineanu, A. Cook, H. Brown-Shaklee, C. DiAntonio, D. Kammler, A. Roach, Sintering processes in direct ink write additive manufacturing: A mesoscopic modeling approach, Acta Materialia, 169 (2019) 60-75.

[31] B. Dzepina, D. Balint, D. Dini, A phase field model of pressure-assisted sintering, Journal of the European Ceramic Society, 39 (2019) 173-182.

[32] J. Hötzer, M. Seiz, M. Kellner, W. Rheinheimer, B. Nestler, Phase-field simulation of solid state sintering, Acta Materialia, 164 (2019) 184-195.

[33] Q. Yang, A. Kirshtein, Y. Ji, C. Liu, J. Shen, L.Q. Chen, A thermodynamically consistent phase-field model for viscous sintering, Journal of the American Ceramic Society, 102 (2019) 674-685.

[34] Y. Yang, O. Ragnvaldsen, Y. Bai, M. Yi, B.-X. Xu, 3D non-isothermal phase-field simulation of microstructure evolution during selective laser sintering, npj Computational Materials, 5 (2019) 1-12.

[35] I. Greenquist, M. Tonks, M. Cooper, D. Andersson, Y. Zhang, Grand potential sintering simulations of doped UO2 accident-tolerant fuel concepts, Journal of Nuclear Materials, 532 (2020) 152052.

[36] I. Greenquist, M.R. Tonks, L.K. Aagesen, Y. Zhang, Development of a microstructural grand potential-based sintering model, Computational Materials Science, 172 (2020) 109288.

[37] W. Yan, W. Ma, Y. Shen, Powder sintering mechanisms during the pre-heating procedure of electron beam additive manufacturing, Materials Today Communications, 25 (2020) 101579.

[38] Z. Zhang, X. Yao, P. Ge, Phase-field-model-based analysis of the effects of powder particle on porosities and densities in selective laser sintering additive manufacturing, International Journal of Mechanical Sciences, 166 (2020) 105230.

[39] V. Ivannikov, F. Thomsen, T. Ebel, R. Willumeit-Römer, Capturing shrinkage and neck growth with phase field simulations of the solid state sintering, Modelling and Simulation in Materials Science and Engineering, 29 (2021) 075008.

[40] R. Termuhlen, X. Chatzistavrou, J.D. Nicholas, H.-C. Yu, Three-dimensional phase field sintering simulations accounting for the rigid-body motion of individual grains, Computational Materials Science, 186 (2021) 109963.

[41] X. Wang, Y. Liu, L. Li, C.O. Yenusah, Y. Xiao, L. Chen, Multi-scale phase-field modeling of layer-by-layer powder compact densification during solid-state direct metal laser sintering, Materials & Design, 203 (2021) 109615.

[42] Z. Zhao, X. Zhang, H. Zhang, H. Tang, Y. Liang, Numerical investigation into pressure-assisted sintering using fully coupled mechano-diffusional phase-field model, International Journal of Solids and Structures, 234 (2022) 111253.

[43] M. Seiz, H. Hierl, B. Nestler, An improved grand-potential phase-field model of solid-state sintering for many particles, Modelling and Simulation in Materials Science and Engineering, 31 (2023) 055006.

[44] M. Seiz, H. Hierl, B. Nestler, Unravelling densification during sintering by multiscale modelling of grain motion, Journal of Materials Science, 58 (2023) 14051-14071.

[45] M. Seiz, H. Hierl, B. Nestler, W. Rheinheimer, Revealing process and material parameter effects on densification via phase-field studies, Scientific Reports, 14 (2024) 5350.

[46] P. Munch, V. Ivannikov, C. Cyron, M. Kronbichler, On the construction of an efficient finite-element solver for phase-field simulations of many-particle solid-state-sintering processes, Computational Materials Science, 231 (2024) 112589.

[47] M. Yi, W. Wang, M. Xue, Q. Gong, B.-X. Xu, Modeling and simulation of sintering process across scales, Archives of Computational Methods in Engineering, 30 (2023) 3325-3358.

[48] J. Pan, Modelling sintering at different length scales, International Materials Reviews, 48 (2003) 69-85.

[49] E.A. Olevsky, V. Tikare, T. Garino, Multi-scale study of sintering: a review, Journal of the American Ceramic Society, 89 (2006) 1914-1922.

[50] Q. Yang, Y. Gao, A. Kirshtein, Q. Zhen, C. Liu, A free-energy-based and interfacially consistent phase-field model for solid-state sintering without artificial void generation, Computational Materials Science, 229 (2023) 112387.

[51] K. Shinagawa, Simulation of grain growth and sintering process by combined phase-field/discrete-element method, Acta materialia, 66 (2014) 360-369.

[52] K. Ahmed, C.A. Yablinsky, A. Schulte, T. Allen, A. El-Azab, Phase field modeling of the effect of porosity on grain growth kinetics in polycrystalline ceramics, Modelling and Simulation in Materials Science and Engineering, 21 (2013) 065005.

[53] T.D. Oyedeji, Y. Yang, H. Egger, B.-X. Xu, Variational quantitative phase-field modeling of nonisothermal sintering process, Physical Review E, 108 (2023) 025301.

[54] C. Liang, Y. Yin, W. Wang, M. Yi, A thermodynamically consistent non-isothermal phase-field model for selective laser sintering, International Journal of Mechanical Sciences, 259 (2023) 108602.

[55] A. Ishii, K. Kondo, A. Yamamoto, A. Yamanaka, Phase-field modeling of solid-state sintering with interfacial anisotropy, Materials Today Communications, 35 (2023) 106061.

[56] P. Barai, T. Kinnibrugh, M. Wolfman, J. Garcia, X. Wang, T.T. Fister, H. Iddir, V. Srinivasan, Phase Field Modeling of Pressure Induced Densification in Solid Electrolytes, JOM, (2024) 1-12.

[57] K. Ahmed, C. Yablinsky, A. Schulte, T. Allen, A. El-Azab, Phase field modeling of the effect of porosity on grain growth kinetics in polycrystalline ceramics, Modelling and Simulation in Materials Science and Engineering, 21 (2013) 065005.

[58] H. Ding, P.D. Spelt, C. Shu, Diffuse interface model for incompressible two-phase flows with large density ratios, Journal of Computational Physics, 226 (2007) 2078-2095.

[59] F. Boyer, A theoretical and numerical model for the study of incompressible mixture flows, Computers & fluids, 31 (2002) 41-68.

[60] H. Abels, H. Garcke, G. Grün, Thermodynamically consistent, frame indifferent diffuse interface models for incompressible two-phase flows with different densities, Mathematical Models and Methods in Applied Sciences, 22 (2012) 1150013.

[61] X. Yang, Y. Gong, J. Li, J. Zhao, Q. Wang, On hydrodynamic phase field models for binary fluid mixtures, Theoretical and Computational Fluid Dynamics, 32 (2018) 537-560.

[62] E.A. Olevsky, Theory of sintering: from discrete to continuum, Materials Science and Engineering: R: Reports, 23 (1998) 41-100.

[63] P.C. Millett, M.R. Tonks, S. Biner, L. Zhang, K. Chockalingam, Y. Zhang, Phase-field simulation of intergranular bubble growth and percolation in bicrystals, Journal of Nuclear Materials, 425 (2012) 130-135.

[64] H. Riedel, H. Zipse, J. Svoboda, Equilibrium pore surfaces, sintering stresses and constitutive equations for the intermediate and late stages of sintering—II. Diffusional densification and creep, Acta metallurgica et materialia, 42 (1994) 445-452.

[65] H. Djohari, J.J. Derby, Transport mechanisms and densification during sintering: II. Grain boundaries, Chemical Engineering Science, 64 (2009) 3810-3816.

[66] H. Djohari, J.I. Martínez-Herrera, J.J. Derby, Transport mechanisms and densification during sintering: I. Viscous flow versus vacancy diffusion, Chemical Engineering Science, 64 (2009) 3799-3809.

[67] M. Doi, S.F. Edwards, S.F. Edwards, The theory of polymer dynamics, oxford university press, 1988.